\documentstyle[sprocl,epsf]{article}

\def\Journal#1#2#3#4{{#1} {\bf #2}, #3 (#4)}

\def\NPB{{\em Nucl. Phys.} B}
\def\PLB{{\em Phys. Lett.} B}
\def\PRL{\em Phys. Rev. Lett.}
\def\PRD{{\em Phys. Rev.} D}

\def\d{\partial}
\def\bV{\bar{V}}
\def\bM{\bar{M}}
\def\bm{\bar{m}}
\def\blam{\bar{\lambda}}
\def\be{\begin{equation}}
\def\ee{\end{equation}}
\def\bea{\begin{eqnarray}}
\def\eea{\end{eqnarray}}

\begin{document}

\title{MASSIVE $\phi^4$ MODEL AT FINITE TEMPERATURE \\
 -- RESUMMATION PROCEDURE {\it A LA} RG IMPROVEMENT --}

\author{HISAO NAKKAGAWA$^{*}$ AND HIROSHI YOKOTA$^{**}$}

\address{Institute for Natural Science, Nara University, 1500 Misasagi-cho, \\
     Nara 631-8502, JAPAN \\
     $^{*}\,$nakk@daibutsu.nara-u.ac.jp, 
             $^{**}\,$yokotah@daibutsu.nara-u.ac.jp} 

\maketitle\abstracts{In this paper the phase structure of the massive
$\lambda \phi^4$ model at finite temperature ($T \neq 0$) is
investigated by applying a resummation method inspired by the
renormalization-group (RG) improvement to the one-loop effective potential.
The resummation method {\it a la} RG-improvement is shown
to work quite succesfully by resumming up systematically large
correction-terms of $O(\lambda T/\mu)$ and of $O(\lambda (T/\mu)^2)$.
The temperature-dependent phase transition of the model is shown to proceed
through the second order transition. The critical exponents are determined analytically and are compared with those in other analyses.}

\section{Introduction}
Understanding the phase structure and the mechanism of phase transition of
quantum field theories at finite temperature/density is important to
understand the evolution of Universe and the physics to be searched
by the ultrarelativistic
heavy-ion experiments planned at the BNL-RHIC and CERN-LHC~\cite{QM97}.
  
To investigate analytically the phase structure of relativistic quantum
field theory, the effective potential (EP) is used as a powerful tool.
Perturbative calculation of the EP at finite temperature, however,
suffers from various troubles: the poor convergence or the breakdown of
the loop expansion~\cite{Linde}, and the strong dependence on the
renormalization-scheme (RS). These troubles have essentially the same origin,
i.e., they come about with the emergence of large perturbative corrections
depending explicitly on the RS. Thus to break a way out of these troubles
we must carry out the systematic
resummation~\cite{Braaten,Dolan,Braaten2,Camelia2} of dominant large
correction terms, and at the same time we must also solve the
problem of the RS-dependence.

Recently simple but very efficient renormalization group (RG)
improvement procedures for resumming dominant large correction terms are
proposed in vacuum~\cite{Bando} and in thermal~\cite{NY1} field theories.
This procedure was originally proposed to solve the problem of the strong
RS-dependence of the EP calculated through the loop-expansion method.

It is worth noticing here that in the massive scalar $\lambda \phi^4$ model at
high temperature the large correction terms appearing in the $L$-loop EP have
the structures as follows; i) terms proportional to powers of the
temperature $T$: i-a) $(\lambda (T/\mu)^2)^L$, i-b) $(\lambda (T/\mu))^L$,
and ii) terms proportional to powers of the $logs$:
ii-a) $(\lambda \ln (T/\mu))^L$, ii-b) $(\lambda \ln (M/\mu))^L$,
where $M$ is the large mass scale appearing in the theory.

In this paper we present the result of application of the
resummation procedure {\it a la} RG~\cite{NY1,NY2} to the massive scalar
$\phi^4$ model renormalized at the temperature of the environment $T$.
We have found that the proposed resummation procedure {\it a la} RG works
efficiently, not only by resolving the problem of the RS-dependence, but
also by properly as well as systematically resumming terms having the
structures i-a) and i-b) above. As for the details of the analyses, 
see Ref.~9 and the paper to appear~\cite{NY3}.

\section{Improving the effective potential through resummation and the
phase structure of massive $\phi^4$ model at $T \neq 0$}
Let us consider the massive scalar $\phi^4$ model at finite temperature
renormalized at an arbitrary mass-scale $\mu$ and at the temperature of the
environment $T$ (hereafter we call this scheme as the $T$-renomalization).
The key idea to resolve the RS-ambiguity is to use correctly and efficiently
the fact that the exact EP satisfies a homogeneous renormalization group
equation (RGE) with respect to change of the arbitrary parameter
$\mu \to \bar{\mu}=\mu e^{t}$.

In the scalar $\phi^4$ model the dominant large corrections appear as a power
function of the effective variable $\tau$ (for more details, see
Refs.~9 and 10)
\bea
 \tau/\lambda &\equiv& \Delta_1 \nonumber \\ 
 & = & \displaystyle{
             \frac{T^2}{2 \pi^2M^2} \left\{ L_1 \left( \frac{T^2}{M^2} \right)
               - \frac{\pi^2}{12} \right\}  
             - \frac{1}{2 \pi^2} L_2 \left(
               \frac{T^2}{\mu^2} \right)
               + \frac{b}{2} \left( \ln \frac{M^2}{\mu^2} -1 \right)},
\eea
where $b=1/16\pi^2$, $M^2 = m^2 + \lambda \phi^2/2$ and $M^2 \Delta_1$ is
nothing but (a part of) the renormalized one-loop self-energy correction,
\bea
 M^2 \Delta_1 &\equiv& \frac12 \sum\!\!\!\!\!\!\!\int \frac{1}{k^2-M^2} +
       \mbox{(one-loop counter term)}, \\ 
    L_i \left( \frac{1}{a^2} \right) &\equiv& \frac{\d^i}{\d (a^2)^i}
        L_0 \left( \frac{1}{a^2} \right) \ , \ \ (i \ge 1), \nonumber \\
    L_0 \left( \frac{1}{a^2} \right) &\equiv&
       \int_0^{\infty} k^2 \, dk \, \ln [ 1 - \exp \{ - \sqrt{k^2+a^2} \} ]
       \ . \nonumber
\eea

The resummation of dominant $O(\lambda (T/\mu)^2)$ terms in the
$T$-renomalization can be automatically performed through renormalization,
giving the remormalized mass-squared
$m^2 \simeq m_0^2 + \frac{1}{24}\lambda T^2$, appearing as a mass-term in the
propagator with which the perturbative calculation is performed, where $m_0$
denotes the renormalized mass in the vacuum theory.

At the one-loop level the RGE's satisfied by the renormalized coupling and
mass-squared can be solved exactly, giving solutions to the running parameters
$\bm^2$ and $\blam$ as
\bea
 & & \bM^2 = \bm^2 + \frac12 \blam \phi^2 , \ \ \bm^2 = m^2 f^{-1/3} , \ \
 \blam = \lambda f^{-1} ,  \\
 & & f = 1 - 3 \lambda \left[ bt + \frac{1}{2 \pi^2} \left\{ L_2 
             \left(\frac{T^2}{\bar{\mu}^2}\right) - L_2
             \left(\frac{T^2}{\mu^2}\right) \right\} \right]. \nonumber
\eea

Up to now $\bar{\mu}$ in the above eqations (3) can be arbitrary, with 
$\mu$ being fixed at the initial value of renormalization. {\it Our
RG-improvement procedure, i.e., the resummation procedure {\it a la} RG, can
be carried out by choosing the RS-fixing parameter $\bar{\mu}$ so as to
satisfy $\bar{\tau}(t)=0$, namely to make the one-loop radiative correction
to the mass fully vanish.}

The RS-fixing equation $\bar{\tau}(t)=0$ gives the mass-gap
equation~\cite{Camelia2,NY1,NY2}
\be
   M^2 = m^2 + f(\bM^2)\bM^2 - f(\bM^2)^{2/3} m^2  \ ,
\ee
which determines, in the HT regime where $T/\mu \gg 1$, the RS-parameter
$\bar{\mu}$ being exact up to $T$-independent constant as
\be
 \bar{\mu} = \frac{\bM}{2} \ . 
\ee

Now we can study the consequences of the RG-improvement in the
$T$-renormalization, with solutions $\blam$, $\bm^2$, and $\bar{\mu}$, Eqs.~(3)
and (5). The RG improvement can then be performed
analytically~\cite{NY1,NY2,NY3}, obtaining the improved EP as
\bea
 \bV_1 &=& \frac12 \bm^2 \phi^2 + \frac{1}{4!} \blam \phi^4 + \bar{h}\bm^4
             \nonumber \\
       & &   + \frac{\bM^4}{2} \left[ -\frac{b}{4}
             + \frac{T^4}{\pi^2 \bM^4} L_0 \left( \frac{T^2}{\bM^2} \right) 
             - \frac{T^2}{2\pi^2 \bM^2} L_1 \left( \frac{T^2}{\bM^2} \right)
             - \frac{T^2}{24 \bM^2} \right] \\ 
    &=& \frac12 \bm^2 \phi^2 + \frac{1}{4!} \blam \phi^4 - \frac{\bm^4}{2 \blam}
             - \frac{T \bM^3}{48 \pi} + \cdots.
\eea

With the RG-improved EP, $\bV_1$, Eqs.~(6) and (7), we can see the nature of
the temperaure-dependent phase-transition of the model;
i) At low temperature below $T_c \sim \sqrt{24|m^2|/\lambda}$ the EP has
twofold structure showing the existence of two phases, Fig.~1a, the ordinary
mass phase and the small mass phase. The ordinary mass phase, with its
counterpart in the tree-level potential, is the symmetry-broken phase
which develops its minimum at
$\phi=\phi_0 \sim \{T_c^4|m^2|^3/\lambda\}^{1/10}$. The small mass phase is
a new ``symmetric'' phase, without having any counterpart in the tree-level
potential, with a linearly decreasing potential unbounded from below,
indicating the simple $\phi^4$ model becoming an unstable theory at
low temperature. As the temperature becomes higher the the minimum of the
ordinary mass phase eventually diminishes, and ii) at the critical temperature
$T_c$ the minimum of the potential at non-zero $\phi$ completely disappears.
The EP shows a symmetric structure in $\phi$ with the minimum at
$\phi=0$, $V(\phi)-V(0) \propto \phi^{\delta+1}$, $\delta \sim 5.0$, 
Fig.~1b, and iii) at high temperature above $T_c$ the EP remains symmetric
in $\phi$ with its minimum at $\phi=0$. Transition between the
symmetry-broken phase at low temperature and the symmetric phase at high
temperature proceeds through the second order transition. 

\begin{figure}[t]\epsfxsize=11cm
\centerline{\epsfbox{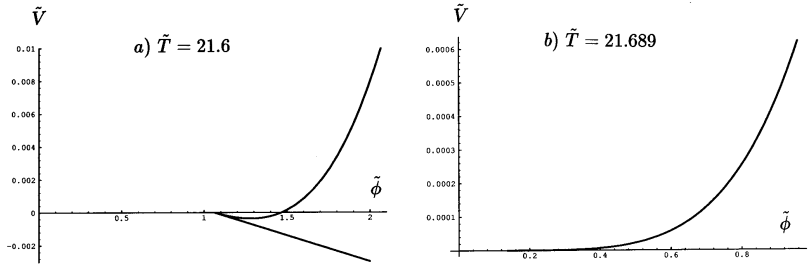}}
\vskip4mm
\caption{The RG improved effective potential of the massive $\phi^4$
      model at two temperatures in the $T$-renormalization: 
      a) $\tilde{T}_1=21.6$ and b) $\tilde{T}_2=\tilde{T}_c=21.689$
      with $\tilde{T} \equiv T/|m_0|$.
      $\tilde{V} \equiv [ \bV_1 (\tilde{\phi}) - \bV_1 (\tilde{\phi}_{min}) ]/
      |m_0|^4$, $\tilde{\phi} \equiv \phi/|m_0|$ and $\lambda = 1/20$.}
\label{fig1}
\end{figure}

\section{Critical exponents} 
Here we present the critical exponents determined from the RG-improved
one-loop EP in the $T$-renormalization, Eqs.~(6) and (7).
In this case we can calculate the critical exponents through analytic manipulation.

The definition of the critical exponents are as follows;
1) On the behavior at $\phi=\phi_0$ around $T \simeq T_c$:
$\phi_0 \propto (T_c- T)^{\beta}$,
$d^2V/d \phi^2 |_{\phi=\phi_0} \propto | T_c - T|^{\gamma}$,
$V(\phi_0)-V(0) \propto |T_c-T|^{2-\alpha}$.
2) On the behavior at $T = T_c$:
$V(\phi)-V(\phi_0=0) \propto \phi^{\delta+1}$ or 
$dV/d \phi \propto \phi^{\delta}$.
Here $\phi_0$ denotes the position of the true minimum, and $T_C$
denotes the critical temperature, which are determined as
$\phi_0 \simeq \{ 54^3 T_c^4 |m^2|^3/\lambda (8 \pi)^4 \}^{1/10}$,
$T_c \simeq \sqrt{24|m_0^2|/\lambda} - 3 |m_0^2|/ 4 \pi \mu $.

Results are summarized in Table~1, showing that our result deviates
significantly from the mean-field values and agrees reasonably with
the experimental data~\cite{expri,lattice}.
\begin{table}[t]
\caption{Critical exponents obtained from various methods.}
\vspace{0.2cm}
\begin{center}
\footnotesize
\begin{tabular}{|l|c|c|c|c|}
\hline
   {}&\raisebox{0pt}[13pt][7pt]{$\beta$} & $\gamma$ & $\delta$ & $\alpha$ \\ \hline
  \raisebox{0pt}[13pt][7pt]{Our result} & 0.3  & 1.2  & 5.0  & 0.2 \\ \hline
  \raisebox{0pt}[13pt][7pt]{mean-field} & 0.5  & 1.0  & 3.0  & 0.0 \\ \hline
  \raisebox{0pt}[13pt][7pt]{lattice~\cite{lattice}} & 0.324 & 1.24 & 4.83
           & 0.113 \\ \hline
  \raisebox{0pt}[13pt][7pt]{experimental~\cite{expri}} & 0.325 & 1.24 & {}
  & ~0.112~ \\ \hline
\end{tabular}
\end{center}
\end{table}

\section{Summary and discussion}
In this paper we proposed a new resummation method inspired by the
renor\-malization-group improvement. Applying this resummation procedure
{\it a la} RG-improvement to the one-loop effective potential in the massive
scalar $\phi^4$ model renormalized at the temperature of the environment $T$,
we found important observations;  The $O(\lambda (T/\mu)^2)$-term
resummation, thus the so-called hard-thermal-loop resummation~\cite{Braaten}
in this model, can be simply done through the $T$-renormalization itself.
With the lack of freedom we can set only one condition to choose the RS-fixing
parameter, which actually ensures to absorb the large terms of
$O(\lambda T/\mu)$, thus only the partial resummation of these terms can be
carried out.

It is to be noted that all the results obtained are essentially
the same as those in the $T_0$-renormalization case~\cite{NY2,NY3}:
the second order phase transition between the ordinary mass broken phase at
low temperature and the symmetric phase at high temperature, and the existence
of the unstable small mass phase at low temperature. In this sense our
resummation method gives stable results  so long as the terms of
$O(\lambda T/\mu)$ are systematically resummed. The critical exponents are
determined by analytic manipulation, showing the
significant deviation from the mean-field values and the reasonable
agreement with the experimental data~\cite{expri,lattice}.
For details, see Refs.~9 and 10.

As noticed above, the RG-improved EP in the simple massive $\phi^4$
model has, below the critical temperature $T_c$, an unstable small-mass phase
in addition to the ordinary symmetry-broken phase. This unstable
phase also appears in the same model at exact zero-temperature,
indicating its appearence being not the artifact due to the crudeness of
approximation on the temperature-dependent correction terms. Though
the origin of its appearence is not fully understood, it may have a relation
with the triviality of the model, which is an interesting problem for further
studies. The $O(N)$ symmetric model in the large-$N$ limit exists as a
stable theory without having such an unstable phase~\cite{NY1}.

\section*{Acknowledgments}
This work is partly supported by the Special Research Grant-in-Aid
of the Nara University.

\end{document}